\documentstyle[12pt]{article}
\begin{document}

\begin{center}
International Journal of Bifurcation and Chaos, Vol. {\bf 9}, No. 3 (1999)
533-539

\vspace{2cm} {\bf SYNCHRONIZATION OF CHAOTIC SYSTEMS DRIVEN BY IDENTICAL
NOISE }\\[\baselineskip]

\thispagestyle{empty}

{\ }{\small B. Kaulakys}$^1${\small , F. Ivanauskas}$^2${\small \ and T.
Me\v skauskas}$^2${\small \\[\baselineskip] \vspace{-0.5\baselineskip} }$^1$
{\small Institute of Theoretical Physics and Astronomy, A. Go\v stauto 12,
2600 Vilnius, Lithuania\\[\baselineskip]\vspace{-0.5\baselineskip}and
Department of Physics, Vilnius University, Saul\.etekio al. 9, 2040 Vilnius,
Lithuania\\[\baselineskip]}\vspace{-0.5\baselineskip}$^2${\small Department
of Mathematics, Vilnius University, Naugarduko 24, 2006 Vilnius, Lithuania}
\\[\baselineskip]
\end{center}

\baselineskip=1.5\normalbaselineskip
\vspace{2cm}

\noindent {\it Abstract}: An analysis of transition from chaotic to
nonchaotic behavior and synchronization in an ensemble of systems driven by
identical random forces is presented. The synchronization phenomenon is
investigated in the ensemble of particles moving with friction in the
time-dependent potential and driven by the identical noise. The threshold
values of the parameters for transition from chaotic to nonchaotic behavior
are obtained and dependencies of the Lyapunov exponents and power spectral
density of the current of the ensemble of particles on the nonlinearity of
the systems and intensity of the driven force are analyzed.\vspace{1.8cm} \

A shortened version of the paper: \vspace{0.5cm}

\begin{center}
{\bf SYNCHRONIZATION BY IDENTICAL NOISE }\newpage\
\end{center}

\setcounter{page}{1}

\noindent {\bf 1. Introduction}\vspace{0.4cm}

\noindent Often trajectories of the nonlinear dynamical systems are very
sensitive to the initial conditions and unpredictable, i.e. the systems are
chaotic. These systems exhibit an apparent random behavior. It might be
expected that when turning on additional random forces make their behavior
even ''more chaotic''. However, a transition from chaotic to nonchaotic
behavior in an ensemble of particles with different initial conditions
bounded in a fixed external potential and driven by an identical sequence of
random forces was observed by Fahy \& Hamann [1992] and recently analyzed
theoretically and numerically by Kaulakys \& Vektaris [1995a,b] and Chen
[1996]. It has been shown that the ensemble of trajectories in such a case
may become identical at long times. The system becomes not chaotic: the
trajectories are independent on the initial conditions. The similar effects
have been observed in the different systems as well [Yu {\it et al.}, 1990
and Maritan \& Banavar, 1994a] and have resulted to same discussion
concerning the origin and causality of such nonchaotic behavior [Pikovsky,
1994, Maritan \& Banavar, 1994b and Gade \& Basu, 1996].

Moreover, the observed effect resembles a phase transition but does not
depend crucially on the dimension of the space in which the particles move.
This phenomenon has some importance for Monte Carlo simulations and can
influence on the clustering of particles process.

It should be noted that Maritan \& Banavar [1994a] have analyzed the similar
effect using the Langevin equation, however, in the limit in which the time
separation between successive forces become small compared to any
characteristic macroscopic time of the system and two logistic maps linked
with a common noise term.

Here we analyze the similar phenomenon in the ensemble of particles moving
with friction in the time-dependent potential and driven by the discrete
identical noise. We define the threshold values of the parameters for
transition from chaotic to nonchaotic behavior and investigate dependencies
of the Lyapunov exponents and power spectral density on the nonlinearity of
the systems and character of the driven force. Our theoretical analysis is
based on the mapping form of equations of motion for the distance between
the particles and the difference of the velocity of the particles while
numerical calculations are performed according to the derived mapping
equations as well as directly calculating the system's trajectories and the
Lyapunov exponents. The mapping analysis results in the conclusions very
close to those obtained from the direct simulations and numerical
calculations\vspace{1.2cm}

\noindent {\bf 2. Models and Theory}\vspace{0.4cm}

\noindent Consider a system of particles of mass $m$ moving with friction
according to Newton's equations
$$
\frac{d{\bf r}}{dt}={\bf v},\qquad \frac{d{\bf v}}{dt}=-\frac 1m\frac{
dV\left( {\bf r},t\right) }{d{\bf r}}-\gamma {\bf v}\eqno{(1)}
$$
in the time dependent potential $V({\bf r},t)$, e.g. in the potential $
V\left( x,t\right) =x^4-x^2-ax\sin \omega t$, and with the friction
coefficient $\gamma $.

At time intervals $\tau $ the particles are partially stopped and their
velocities are reset to the mixture of some part $\alpha $ of the old
velocities with the random velocity ${\bf v}_k^{ran}$: ${\bf v}^{new}=\alpha
{\bf v}^{old}+{\bf v}_k^{ran}$, where $k$ is the stop number. Note that $
{\bf v}_k^{ran}$ depends on the stop number $k$ but not on the particle. The
simplest and most natural way is to choose the random values of velocity $
{\bf v}_k^{ran}$ from a Maxwell distribution with $k_BT=m=1$, i.e. from the
Gaussian distribution of variance $\sigma ^2=1$. Figure 1 illustrates the
difference of evolution of the ensemble of particles with randomly
distributed (from the Gaussian distribution of variance $\sigma ^2=1$)
initial conditions and perturbed by the replacement ${\bf v}^{new}=\alpha
{\bf v}^{old}+{\bf v}_k^{ran}$ for different values of the time interval $
\tau $ between such perturbations. We see the transition to one (common for
all particles) trajectory for sufficiently small time interval $\tau .$

Theoretically a transition from chaotic to nonchaotic behavior in such a
system may be detected from analysis of the neighboring trajectories of two
particles initially at points ${\bf r}_0$ and ${\bf r}_0^{\prime }$ with
starting velocities ${\bf v}_0$ and ${\bf v}_0^{\prime }$. The convergence
of the two trajectories to the single final trajectory depends on the
evolution with a time of the small variances $\Delta {\bf r}_k={\bf r}
_k^{\prime }-{\bf r}_k$ and $\Delta {\bf v}_k={\bf v}_k^{\prime }-{\bf v}_k$
. From formal solutions ${\bf r}={\bf r}({\bf r}_k,{\bf v}_k,t)$ and ${\bf v}
={\bf v}({\bf r}_k,{\bf v}_k,t)$ of the Newton's equations with initial
conditions ${\bf r}={\bf r}_k$ and ${\bf v}={\bf v}_k$ at $t=0$ it follows
the mapping form of the equations of motion for $\Delta {\bf r}$ and $\Delta
{\bf v}$ [Kaulakys \& Vektaris 1995a,b]:

$$
\pmatrix{\Delta{\bf r}_{k+1}\cr \Delta{\bf v}_{k+1}}=\mbox{\bf T}(\alpha ;
{\bf r}_k,{\bf v}_k,\tau _k)\pmatrix{\Delta{\bf r}_k\cr \Delta{\bf v}_k}
\eqno(2)
$$
where the ${\bf T}$ matrix is of the form
$$
\mbox{\bf T}=
\pmatrix{T_{\bf rr}&\alpha T_{\bf rv}\cr T_{\bf vr}&\alpha
T_{\bf vv}}=
\pmatrix{\displaystyle{\partial{\bf r}\over\partial{\bf r}_k}&\displaystyle{\alpha
{\partial{\bf r}\over\partial{\bf v}_k}}\cr\displaystyle{\partial{\bf v}\over
\partial{\bf r}_k}&\displaystyle{\alpha{\partial{\bf v}\over\partial{\bf v}_k}}}
\eqno(3)
$$
and the time interval $\tau _k$ may depend on the step $k$.

For one-, two- and three-dimensional systems the dimension of the ${\bf T}$
matrix is 2, 4 and 6, respectively.

Matrix elements $T_{{\bf rr}}$ and $T_{{\bf rv}}$ satisfy the equation
$$
{\frac{d^2T_{{\bf r}}}{dt^2}}=\left. -{\frac 1m}\left( T_{{\bf r}}\cdot {\
\frac d{d{\bf r}}}\right) {\frac{dV\left( {\bf r},t\right) }{d{\bf r}}}
\right| _{{\bf r}={\bf r}({\bf r}_k,{\bf v}_k,t)}-\gamma {\frac{dT_{{\bf r}}
}{dt}}\eqno(4)
$$
and initial conditions at $t=0$
$$
T_{{\bf rr}}({\bf r}_k,{\bf v}_k,0)=T_{{\bf vv}}=1,~~~T_{{\bf rv}}=T_{{\bf
vr }}=0
$$
$$
\dot T_{{\bf rr}}({\bf r}_k,{\bf v}_k,0)=\dot T_{{\bf vv}}=0,~\dot T_{{\bf
rv }}=1,~\dot T_{{\bf vr}}=-{\frac 1m}{\frac{d^2V}{d{\bf r}^2}}\Biggl|
_{x=x_k}, \eqno{(5)}
$$
while $T_{{\bf vr}}=\dot T_{{\bf rr}}$ and $T_{{\bf vv}}=\dot T_{{\bf rv}}$.
Here and further the points over the letters express the derivatives with
respect to the time.

Further analysis is based on the general theory of dynamics of classical
systems represented as maps: we can calculate the eigenvalues $\mu _k$ of
the ${\bf T}$ matrix for each step and evaluate the averaged Lyapunov
exponents or $KS$ entropy of the system
$$
\sigma _k=\left\langle {\frac 1{\tau _k}}\ln |\mu _k|\right\rangle =\lim
_{N\to \infty }{\frac 1N}\sum_{k=1}^N{\frac 1{\tau _k}}\ln |\mu _k({\bf r}
_k, {\bf v}_k,{\bf \tau }_k)|\eqno(6)
$$
A criterion for transition to chaotic behavior is
$$
\sigma _{max}=0.\eqno(7)
$$

Comparisons of the threshold values $\tau _c$ for transition to chaos
according to Eqs. (2)-(7) with those from the direct numerical simulations
indicate to the fitness and usefulness of the method (2)-(7) for
investigation of transition from chaotic to nonchaotic behavior in randomly
driven ensemble of systems bounded in the fixed external potential without
the friction [Kaulakys \& Vektaris 1995a,b].\vspace{0.4cm}

\noindent {\bf 3. Results of calculations}\vspace{0.4cm}

\noindent Here we calculate the Lyapunov exponents directly from the
equations of motion and linearized equations for the variances of coordinate
and velocity. Further we extend the same analysis for the systems with
friction in the regularly time depending external field and perturbed by the
identical for all particles random force.

Figures 2-5 represent an extensive analysis of the autonomous system based
on the numerical solutions of the differential equations of motion. Figure 2
shows quite similar behavior of the Lyapunov exponent to that calculated
from the mapping equations of motion (curve (a) of Fig. 2 in the paper
[Kaulakys \& Vektaris 1995a]). Figures 3-5 represent dependences of the
Lyapunov exponents on the different parameters of the model. Areas of the
parameters for which the Lyapunov exponents are negative corresponds to the
nonchaotic Brownian-type motion.

In general, motion in the nonautonomous Duffing potential with friction
describe equations
$$
\dot v=2x-4x^3-\gamma v+a\sin \omega t,\qquad \dot x=v.\eqno{(8)}
$$

For $a=0$ and $\gamma =0$ Eqs. (8) represent motion in the fixed external
potential. As it was mentioned above, in the paper by Kaulakys \& Vektaris
[1995a] some theoretical and numerical analysis of this model was fulfilled
on the bases of the mapping equations (2)-(7).

Figure 6 illustrates evolution of the ensemble of particles with randomly
distributed initial conditions in the nonautonomous Duffing potential with
friction and perturbed by the replacement $v^{new}(k\tau )=\alpha
v^{old}(k\tau )+v_k^{ran},$ $k=1,2,...$. We observe a transition from the
actual chaotic dynamics for large $\tau $ to the nonchaotic common for all
particles trajectory with the decrease of $\tau $. In Fig. 7 we show the
dependence on $\tau $ of the Lyapunov exponents for the motion in the
nonautonomous Duffing potential with friction. For the values of parameters
corresponding to the positive Lyapunov exponents, i.e. without the random
perturbation ($\tau \rightarrow \infty $), the system is chaotic. The
negative Lyapunov exponents for small $\tau $ indicate to the nonchaotic
Brownian-type motion. \vspace{0.4cm}

\noindent {\bf 4. Spectrum of the current noise }\vspace{0.4cm}

\noindent As it has already been observed in the paper by Kaulakys \&
Vektaris [1995b] such systems exhibit the intermittency route to chaos which
provides sufficiently universal mechanism for $1/f$-type noise in the
nonlinear systems. Here we analyze numerically the power spectral density of
the current of the ensemble of particles moving in the closed contour and
perturbed by the common for all particles noise. The simplest equations of
motion for such model are of the form
$$
\dot v=F-\gamma v,\qquad \dot x=v\eqno{(9)}
$$
with the perturbation given by the resets of velocity of all particles after
every time interval $\tau $ according to the identical for all particles
replacement $v^{new}(k\tau )=\alpha v^{old}(k\tau )+v_k^{ran},$ $k=1,2,...$.
For sufficiently small $\tau $ we observe the current power spectral density
$S\left( f\right) $ dependence on the frequency $f$ close to the $1/f$
-dependence (Fig. 8). It should be noted that such spectral density
dependence is nonsensitive to some additional (nonlinear) terms in the
equation for velocity. The essential condition for the $1/f$-type dependence
of the current power spectral density is the random sufficiently strong
perturbation of the particles' velocities (see also Kaulakys \& Me\v
skauskas [1997] for analysis of other systems and different perturbations).

\vspace{0.4cm}

\noindent {\bf 5. Conclusions}\vspace{0.4cm}

\noindent From the fulfilled analysis we may conclude that, first,
synchronization and transition from chaotic to nonchaotic behavior in
ensembles of the identically perturbed by the random force nonlinear systems
may be analyzed as from the mapping form of equations of motion for the
distance between the particles and the difference of the velocity as well as
from the direct calculations of the Lyapunov exponents and, second, the
model of Fahy \& Hamann [1992] may be generalized for the ensemble of
particles moving with friction in the time-dependent potential.

The transition from chaotic to nonchaotic behavior in the ensemble of
particles moving with friction may also be observed in the more realistic
case of motion without the momently stops of the particles. On the other
hand, the motion in the time-dependent potential without the random periodic
perturbations even in the one-dimensional case may be chaotic or nonchaotic,
depending on the system's parameters. Therefore, such generalization of the
model allows to investigate the synchronization phenomenon and transition
from chaotic to nonchaotic behavior effect in an ensemble of systems driven
by identical random forces on the base of relatively simple
(one-dimensional) models for larger variety of the system's dynamics.

Moreover, an ensemble of systems linked with a common external noise may
exhibit the $1/f$-type fluctuations. Our model may easily be generalized for
systems driven by any random forces or fluctuations. On the other hand, the
phenomenon when an ensemble of systems is linked with a common external
noise or fluctuating external fields is quite usual. Thus, an ensemble of
systems in the external random field may provide a sufficiently universal
mechanism of $1/f$--noise.\vspace{0.4cm}

\noindent {\bf Acknowledgment}\vspace{0.4cm}

\noindent The research described in this publication was supported in part
by Grant No. 220 of the Lithuanian State Science and Studies Foundation.
\newpage\

\noindent {\bf References}

\noindent Chen, Y. Y. [1996] ''Why do chaotic orbits converge under a random
velocity resets'', {\it Phys. Rev. Lett}. {\bf 77}(21), 4318 - 4321.

\noindent Fahy, S. \& Hamann, D. R. [1992] ''Transition from chaotic to
nonchaotic behavior in randomly driven systems'', {\it Phys. Rev. Lett.}
{\bf 69}(5), 761-764.

\noindent Gade, P. M. \& Basu, C. [1996] ''The origin of non-chaotic
behavior in identically driven systems'', {\it Phys. Lett}. {\bf A217}(1),
21 - 27.

\noindent Kaulakys, B. \& Vektaris, G. [1995a] ''Transition to nonchaotic
behavior in a Brownian-type motion, {\it Phys. Rev. }{\bf E52}(2),
2091-2094; chao-dyn/9504009.

\noindent Kaulakys, B. \& Vektaris, G. [1995b] ''Transition to nonchaotic
behavior in randomly driven systems: intermittency and 1/f-noise'', {\it
Proc. 13th Int. Conf. Noise in Phys. Syst. and 1/f Fluctuations, }eds.
Bareikis, V. \& Katilius, R. (World Scientific, Singapore) pp.677-680.

\noindent Kaulakys, B. \& T. Me\v skauskas [1997] ''On the 1/f fluctuations
in the nonlinear systems affected by noise, {\it Proc. 14th Int. Conf. Noise
in Phys. Syst. and 1/f Fluctuations, }14 - 18 July 1997, Leuven, Belgium
(World Scientific, Singapore), pp. 126-129; adap-org/9806002;
adap-org/9812003; B. Kaulakys [1999] ''Autoregressive model of 1/f noise''
Phys. Lett. A {\bf 257}(1-2), 37-42.

\noindent Maritan, A. \& Banavar, J. R. [1994a] "Chaos, noise and
synchronization", {\it Phys. Rev. Lett.} {\bf 72}(10), 1451-1454.

\noindent Maritan, A. \& Banavar, J. R. [1994b] "Reply to Comment of
Pikovsky", {\it Phys. Rev. Lett.} {\bf 73}(21), 2932.

\noindent Pikovsky, A. S. [1994] "Comment on "Chaos, noise and
synchronization"", {\it Phys. Rev. Lett.} {\bf 73}(21), 2931.

\noindent Yu, L., Ott. E. \& Chen, Q. [1990] ''Transition to chaos for
random dynamical systems'', {\it Phys. Rev. Lett.} {\bf 65}(24), 2935-2938.

\newpage\

\begin{center}
Figure captions for the paper
\end{center}

\noindent Fig. 1. Illustration of the difference of evolution in the $\left(
x,v\right) $ -space of the ensemble of particles in the autonomous Duffing
potential $V(x)=x^4-x^2$ with randomly distributed initial conditions and
perturbed at time intervals $\tau $ by the replacement $v^{new}(k\tau
)=\alpha v^{old}(k\tau )+v_k^{ran},$ $k=1,2,...$with $\alpha =0.5$. For the
relatively large $\tau =2$ there is no transition to the common trajectory,
for smaller $\tau =0.8$ the clustering process of particles with different
initial conditions is relatively slow while for sufficiently small $\tau
=0.6 $ a collapse to the common trajectory at the time moment $t=100$ is
evident.\vspace{0.3cm}

\noindent Fig. 2. Lyapunov exponent (multiplied by $\tau )$ from the direct
calculations vs the time $\tau $ between the resets of the velocity $
v^{new}(k\tau )=\alpha v^{old}(k\tau )+\beta v_k^{ran},$ $k=1,2,...$ for
motion in the autonomous Duffing potential. \vspace{0.3cm}

\noindent Fig. 3. Lyapunov exponent $\lambda $ for motion in the autonomous
Duffing potential vs the time interval $\tau $ between the resets of the
velocity $v^{new}(k\tau )=\alpha v^{old}(k\tau )+\beta v_k^{ran},$ $
k=1,2,... $ for different values of the parameter $\alpha $. \vspace{0.3cm}

\noindent Fig. 4. As in Fig. 3 but for different values of $\beta $.
\vspace{0.2cm}

\noindent Fig. 5. As in Fig. 3 but for small values of $\beta $.
\vspace{0.3cm}

\noindent Fig. 6. As in the Fig. 1 but for motion according to Eq. (8) in
the nonautonomous Duffing potential with $\gamma =0.07$, $a=5$, $\alpha =0.8$
and $\beta =1$. A transition from the actual chaotic (at $\tau =\infty $) to
the nonchaotic dynamics with the decrease of the time interval $\tau $
between the resets of the velocity is observable. \vspace{0.3cm}

\noindent Fig. 7. Lyapunov exponent $\lambda $ vs the time $\tau $ between
the resets of the velocity $v^{new}(k\tau )=\alpha v^{old}(k\tau )+\beta
v_k^{ran},$ $k=1,2,...$ for different values of the parameter $\alpha $ for
motion in the driven Duffing potential with friction according to Eq.(8).
\vspace{0.3cm}

\noindent Fig. 8. The power spectral density of the current of the ensemble
of particles moving according to Eq. (9) with $F=1$, $\gamma =0.1$ and
perturbed by the common for all particles noise $v^{new}(k\tau )=\alpha
v^{old}(k\tau )+v_k^{ran},$ $k=1,2,...$ with $\alpha =1$ and different
values of $\tau $. The dense lines represent the averaged spectra.

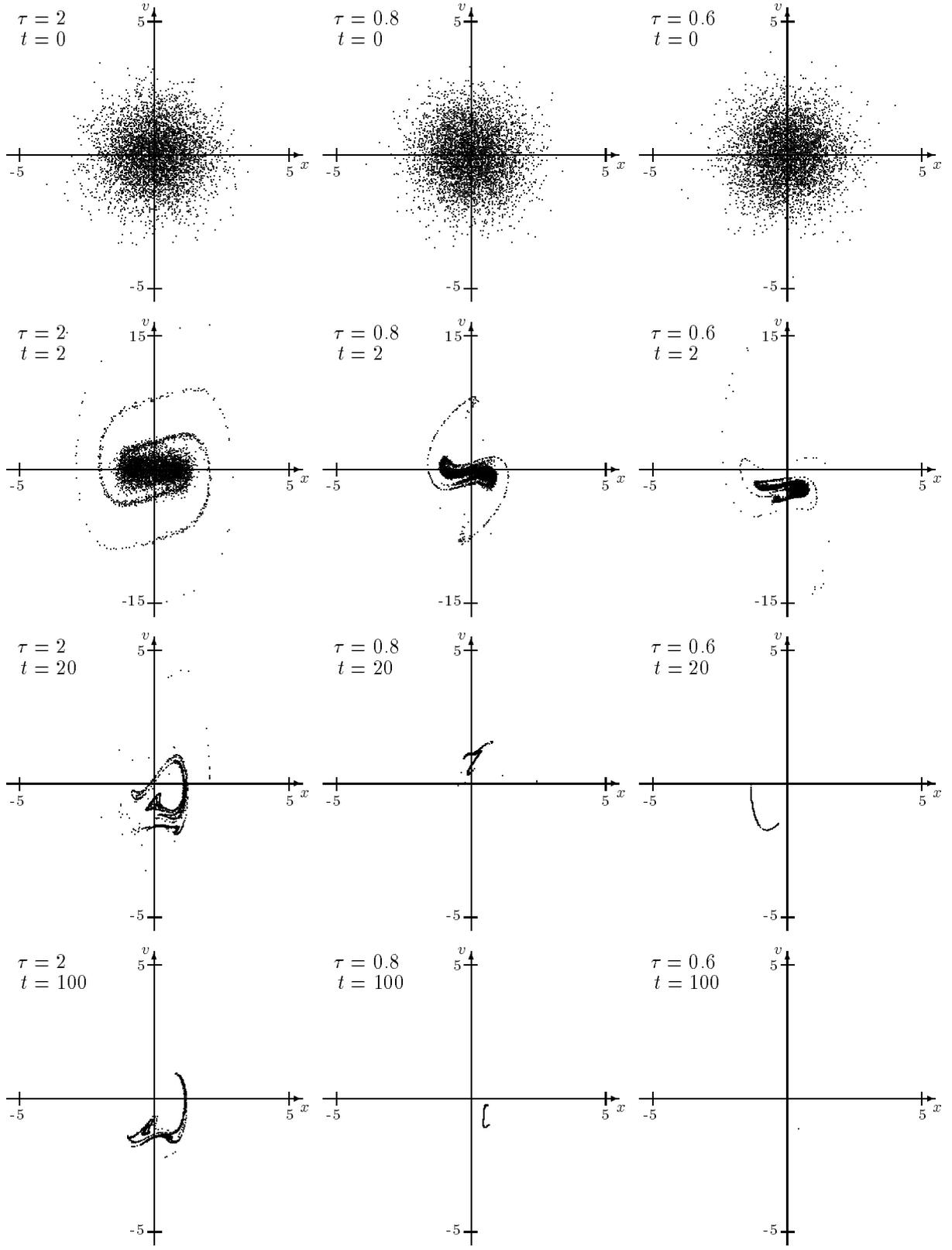
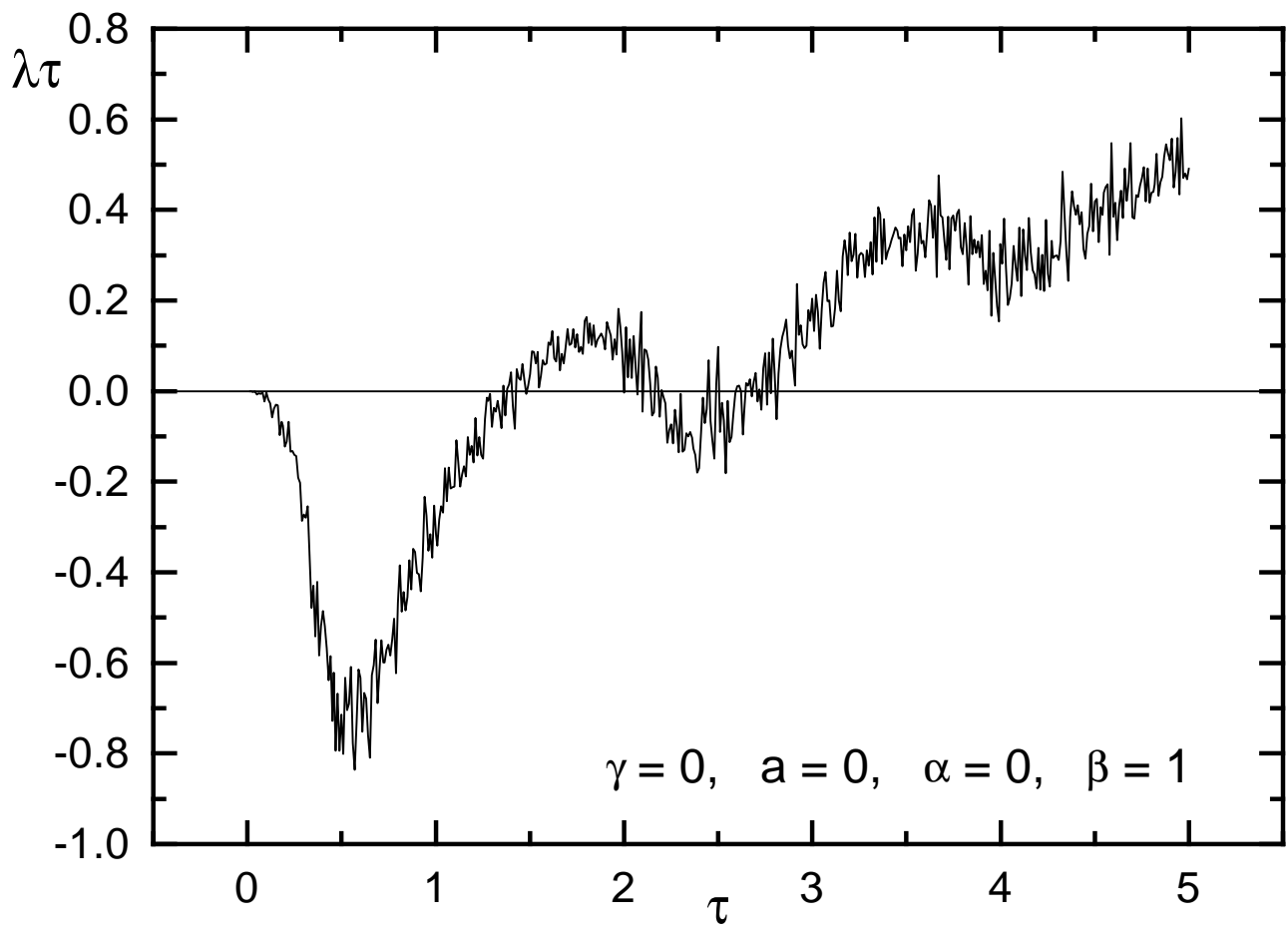
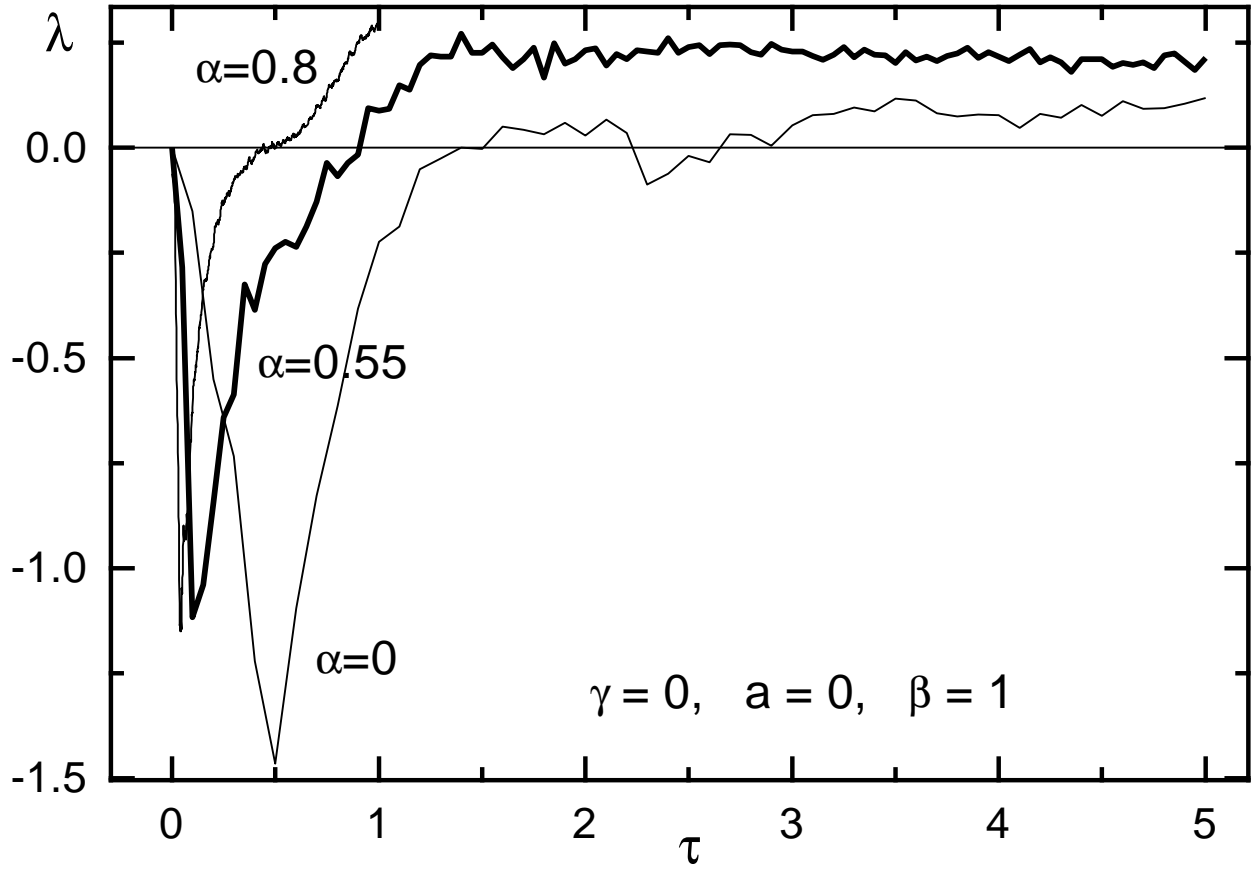
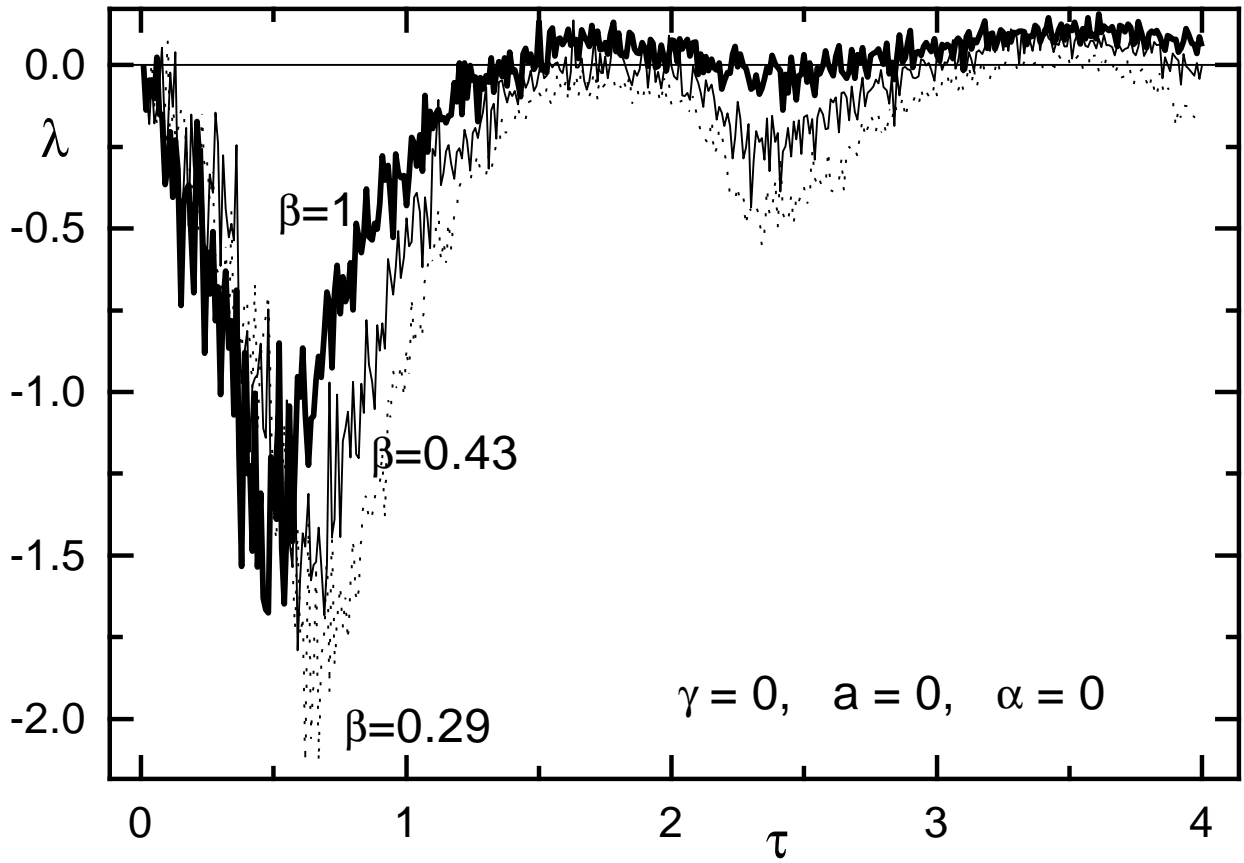
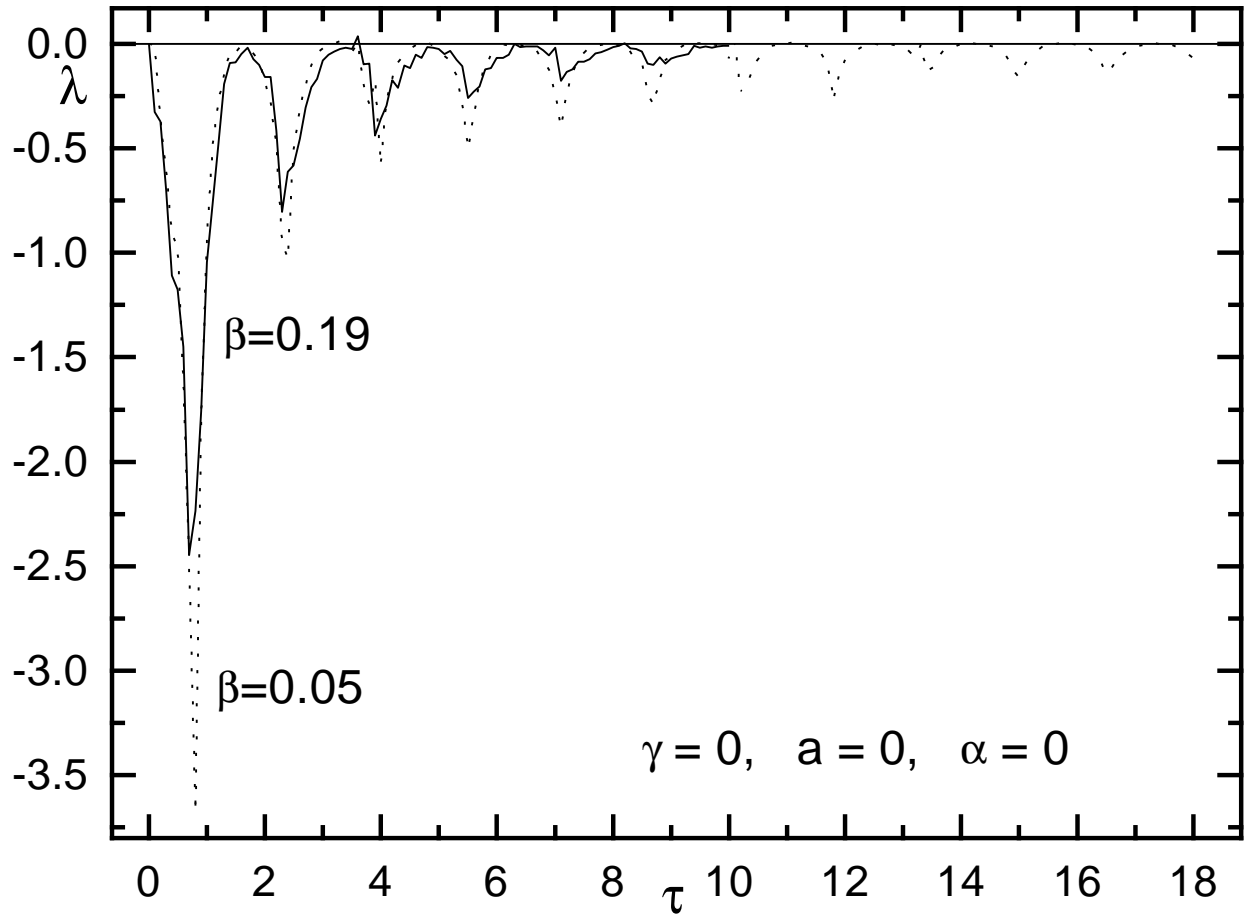
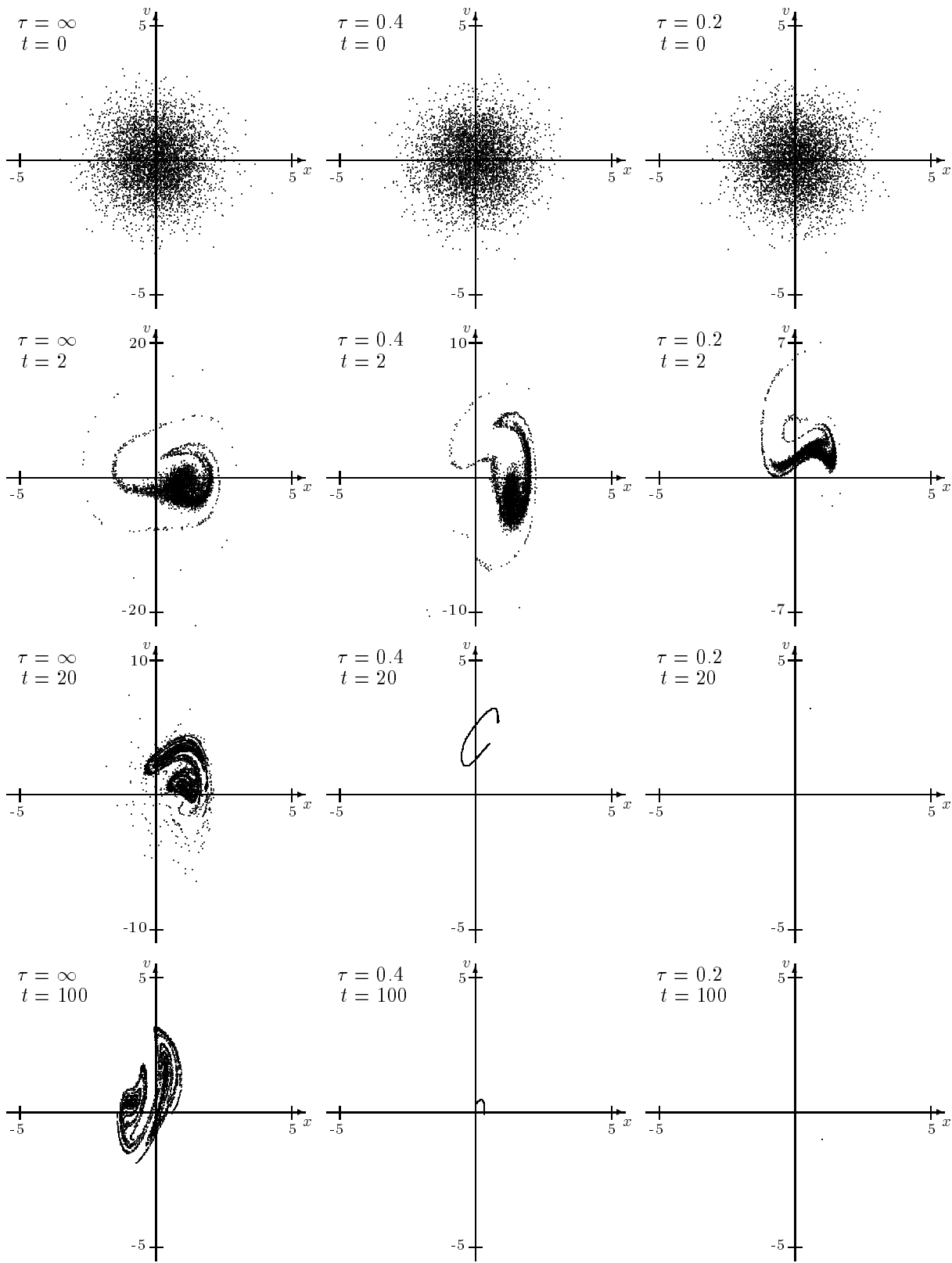
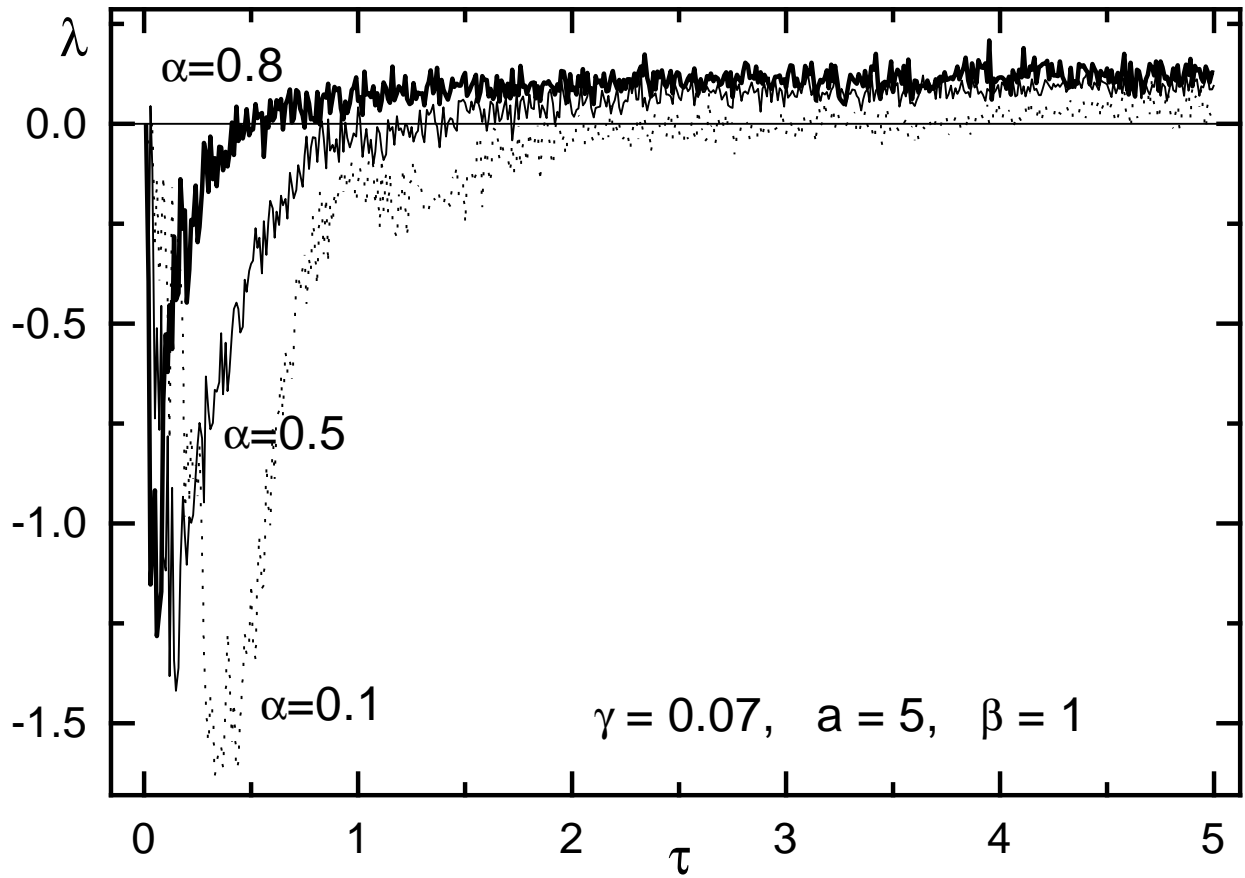
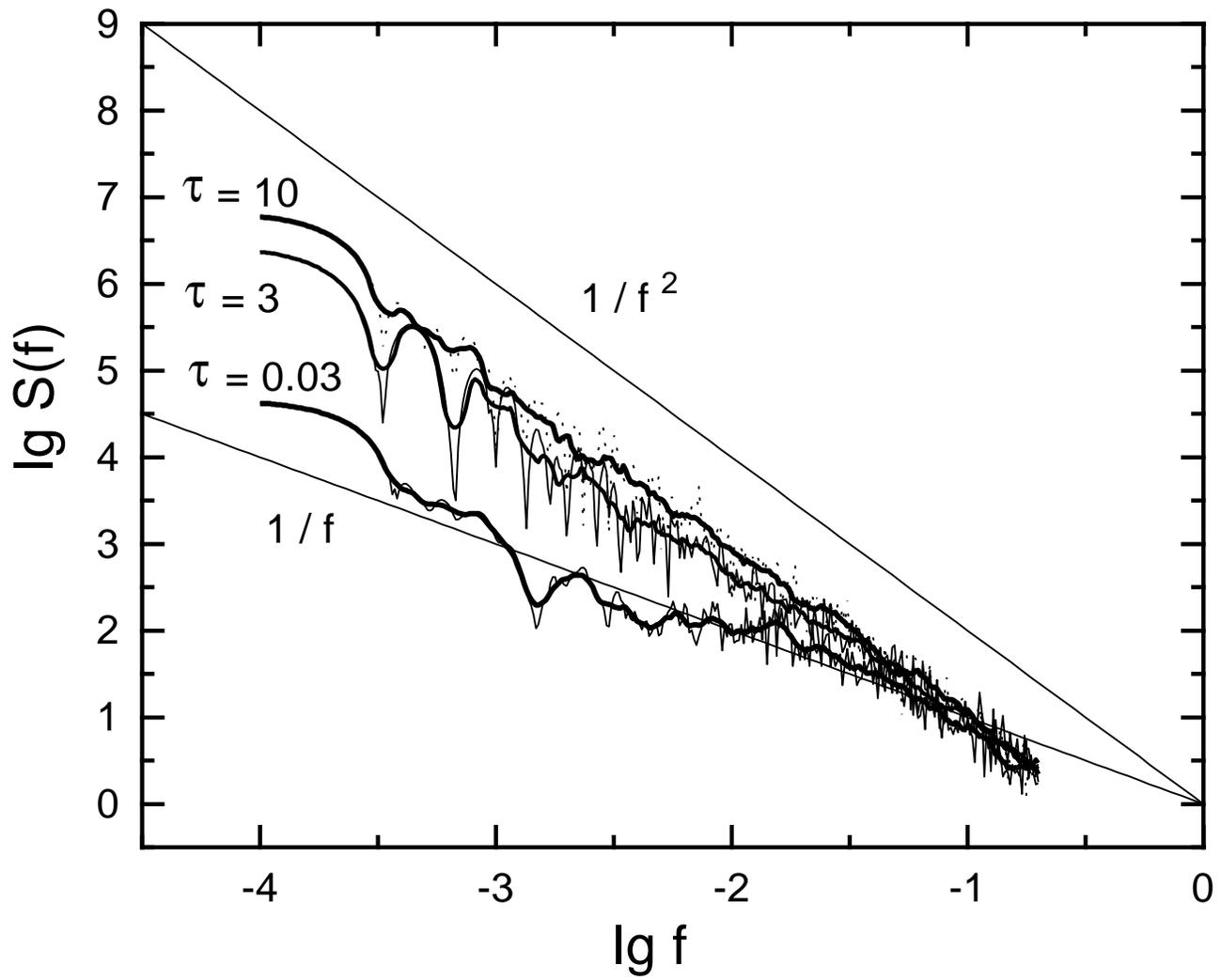
\end{document}